\documentclass[aps,pra,onecolumn,amsmath,amssymb]{revtex4}

\usepackage{amssymb}
\usepackage{graphicx}
\usepackage{dcolumn}
\usepackage{bm}
\usepackage{float}
\begin{document}

\title{Persistent spin squeezing of dissipative Lipkin-Meshkov-Glick Model embedded in a general thermal environment}
\author{Yong-Hong Ma$^{1,2}$}\email{myh_dlut@126.com}
 \author{ Quan-Zhen Ding$^{2}$ }
\author{Ting Yu$^{2}$}\email{ting.yu@stevens.edu}
\affiliation{$^{1}$School of Science, Inner Mongolia University
of Science and Technology, Baotou 014010, P. R. China}
\affiliation{$^{2}$Department of Physics, and Center for Quantum Science and Engineering,
Stevens Institute of Technology, Hoboken, New Jersey 07030, USA}

\date{\today}

\begin{abstract}
We investigate spin squeezing for a Lipkin-Meshkov-Glick (LMG) model coupled to a general non-Markovian environment in a finite temperature regime. Using the non-Markovian quantum state diffusion and master equation approach, we numerically study non-Markovian spin squeezing generation in LMG model.    Our results show that the total spin number $N$, energy $k_{B}T$, and certain coefficients in a LMG model can play a crucial role in generating spin squeezing. In particular, it shows that the maximum spin squeezing can be significantly enhanced when the participating environment has a relatively long memory time.
\end{abstract}

\maketitle

\section{Introduction}
Spin squeezing, as an important quantum resource, has
many potential applications in quantum science and technology such as quantum metrology, atom interferometers \cite{win4,kit,est,gro,rie}, and
quantum information processing due to its close relation
with quantum entanglement \cite{hofm,toth,cava,cava1,rmp}.  A spin squeezing state is defined for an ensemble of spins, whose fluctuation in one collective spin
direction to the mean spin direction is smaller
than the classical limit \cite{kit}. The growing interest of studying the spin squeezing \cite{kit} have arisen mainly from exploring the correlation and many-body entanglement \cite{ren,big,gue} of particles, especially for the purpose of improving the measurement precision \cite{win2,win4,pol,cro,cha,ler}.

As pointed out by Kitagawa and Ueda \cite{kit}, the spin squeezing states can be generated via two main methods: the one-axis twisting (OAT) model and the
two-axis counter-twisting (TACT) model. At present,
a number of methods to generate spin squeezing in atomic systems have been proposed and successfully realized including the transfer of squeezing property from
light to matter and quantum nondemolition measurement of atomic states \cite{app,sch,gro,rie,ema,tas,wmu}. In the past decade, different techniques of preparing high-quality spin-squeezed states have been widely studied in many interesting physical systems including nuclear spins \cite{rud,kork}, NV centers \cite{benn,maprb,maann}, and electron spins \cite{rev}, to name a few.

A realistic analysis of spin squeezing must take into account decoherence and dissipation phenomena. The theoretical descriptions of spin squeezing in the open system context are mostly considered under Born-Markov approximations due to their simplicity in deriving Markov dynamical equation and the interest in the long-time limit. Generally, Markov Langevin equations or the corresponding master equations can be used to characterise the spin squeezing while the environment noises are taken as a weak perturbation and the memory effects are completely ignored \cite{nie,dou,lee,sai,guang,benn,ema,njp2016}.  It is certainly desirable to study the non-Markovian decoherence and robustness of the spin-squeezing generation in a more general environment~\cite{Nori}. The information about squeezing processes in a wide range of time scales including short-time, mid-time and long-time limits are needed in applications where the system of interest is embedded in an environment that exhibits a finite-time correlation. Such non-Markovian features  are crucial for many important physical processes involving strong system-environment interactions or long-range temporal correlations \cite{nonmarkov4,nonmarkov5,nonmarkov6}. Moreover, when the open system is coupled to a structured reservoir, e.g., for quantum
channels that are embedded in physical media such as band-gap crystals \cite{nonmarkov1,nonmarkov2,nonmarkov3}, the environmental memory effects cannot be ignored. In all these cases, the system dynamics can
be substantially different from the Markov ones \cite{nonmarkov3a}.
One of our motivations in this paper is to investigate how spin squeezing survives in the
presence of non-Markovian relaxation on every site based on the LMG model. Our investigations will involve the spin squeezing generation in wide temporal domains and LMG model parameter ranges .

In this paper, we study the dynamics of spin squeezing of a LMG model constituted of $N$ independent spin-1/2 particles coupled to a non-Markovian environment in a finite temperature regime. We derive a non-Markovian master equation by using the quantum state diffusion approach (QSD) \cite{Diosi1,Diosi2,Yudiosi3}. Solving the time evolutions of several operator expectations, we find that open systems exhibiting strong non-Markovian features can substantially modify the dynamics of spin squeezing. More specifically, it shows that the environment memory time can significantly alter the speed of spin squeezing generation. In addition, we show how to choose the effective parameters to achieve the maximum squeezing.

The organisation of material is as follows. In Sec. II we present
the LMG model and give an effective model Hamiltonian. In Sec. III, we derive the master equation based
on the NMQSD equation. In section IV. the
environmental memory effects on spin squeezing are systematically investigated.  We also give a careful description of how to achieve the maximum spin-squeezing by tuning the key parameters of the LGM model. Finally, the conclusion is given in Sec. V. Some details about the equations of motions for the expectations are left to Appendix A.

\section{Hamiltonian of the System} \label{sec:model}
We consider a Lipkin-Meshkov-Glick (LMG) model, which is initially introduced in nuclear physics \cite{LMG} and widely used
in many other fields in the recent years, such as statistical mechanics of quantum spin systems \cite{bot}, Bose-Einstein condensations
\cite{cir}, and superconducting circuits \cite{tso}, those systems may serve as a potential realisation of the model to be discussed in this paper. The Hamiltonian of a LMG model can be described as (setting  $\hbar=1$)
\begin{eqnarray}
\hat{H}_\text{LMG}&=&-\frac{\lambda}{N}\sum_{i<j}
(\sigma_x^i\sigma_x^j+\alpha \sigma_y^i\sigma_y^j)-h\sum_{i}\sigma_z^j\nonumber\\
&&=-\frac{2\lambda}{N}(J_x^2+\alpha J_y^2)-2hJ_z+\frac{\lambda}{2}(1+\alpha),
\end{eqnarray}
where $\sigma_{\beta}$ represents a Pauli matrix, $J_{\beta}=\sum_i\sigma^i_{\beta}/2$ is the collective spin opeartor (denoted as $J_{z}=\frac{1}{2}\sum_{n}(|1\rangle_{n}\langle1|-|0\rangle_{n}\langle0|)$, $J_{x}=\frac{1}{2}\sum_{n}(|1\rangle_{n}\langle0|+|0\rangle_{n}\langle1|)$,
$J_{+}=\sum_{n}|1\rangle_{n}\langle0|$, $J_{-}=\sum_{n}|0\rangle_{n}\langle1|$), and $N$ is the
total spin number. We can obtain several different forms of the LMG Hamiltonian via choosing suitable parameters. Generally, the anisotropic parameter $\alpha$ satisfies $0\leq\alpha\leq 1$. Ma. J \emph{et.al.} have calculated the spin squeezing in the isotropic case ($\alpha=1$) \cite{maj} in a Markov limit. In this paper, we consider the ferromagnetic case ($\lambda>0$) and the isotropic  case simultaneously. Thus, the effective Hamiltonian can be written as
\begin{eqnarray}
\hat{H}&=&-\frac{2\lambda}{N}(J_x^2+J_y^2)-2hJ_z+\lambda \nonumber\\
&=&-\frac{2\lambda}{N}(\textbf{J}^2-J_z^2)-2hJ_z+\lambda.
\end{eqnarray}
where $\textbf{J}$ is the conservation quantity and can be written as $\textbf{J}=\frac{N}{2}(\frac{N}{2}+1)$.
For the sake of simplicity, the effective Hamiltonian can be simplified to the following form
\begin{eqnarray}\label{ham}
\hat{H}_{\rm eff}=aJ_z+bJ_z^2,
\end{eqnarray}
where $a=-2h$ and $b=\frac{2\lambda}{N}$.
\section{Non-markovian master equation at finite temperature}
We start with a LMG model interacting with a bosonic
oscillator environment with total Hamiltonian,
\begin{eqnarray}
 H_{\rm I}=aJ_z+bJ_z^2+\sum_j
\omega_j b^\dag_j b_j+\sum_j (g^*_j
L^\dag b_j +g_j L b^\dag_j),
\end{eqnarray}
where $\omega_{j}, b^{\dagger}_{j}, b_{j}$ are frequencies, creation operators and annihilation
operators of the heat bath respectively, $g_j$
are the coupling constants between the system and the bath, and the Lindblad operator $L=J_-$ here indicates spin damping.
According to Ref. \cite{yu2004}, using the bosonic Bogoliubov transformation, the finite temperature problem is effectively reduced to a zero temperature one. Therefore, the trajectory $\psi_t=|\psi_t(z^*,w^*)\rangle$ satisfies the following NMQSD equation with two independent noises
$z^*_t,w^*_t$ \cite{yu2004}:
\begin{eqnarray}\label{NMQSD1}
\partial_t \psi_t &=& -i H_{\rm sys} \psi_t + Lz^*_t \psi_t -L^\dag\int_0^t ds\;\alpha_1(t,s)
\frac{\delta\psi_t}{\delta z^*_s}\nonumber\\
&&+L^\dag w^*_t \psi_t -L\int_0^t ds\;\alpha_2(t,s)
\frac{\delta\psi_t}{\delta w^*_s},
\end{eqnarray}
where
\begin{eqnarray}
\alpha_1(t,s)
&=& \sum_j (\bar{n}_j+1)|g_j|^2
e^{-i\omega_j(t-s)},\nonumber\\
\alpha_2(t,s) &=& \sum_j \bar{n}_j |g_j|^2
e^{i\omega_j(t-s)},
\end{eqnarray}
 are the bath correlation functions.  $z_t^*=-i\sum_j\sqrt{\bar{n}_j+1}\,\,g^*_j z_j^*
e^{i\omega_j t}$ and $w_t^*=-i\sum_j\sqrt{\bar{n}_j}\,\,g^*_j w_j^*
e^{-i\omega_j t}$ denote  two independent and complex
Gaussian noises satisfying ${\mathcal M}[z_t]={\mathcal
M}[z_t z_s]=0,\,\,{\mathcal M}[z_t^* z_s] = \alpha_1(t,s),
{\mathcal M}[w_t]={\mathcal M}[w_t w_s]=0$ and ${\mathcal
M}[w_t^*w_s] = \alpha_2(t,s)$.
Note here ${\mathcal M}[\,\cdot\,]=\int\frac{dz^2}{\pi}e^{-|z|^2}[\,\cdot\,]$
describes the statistical mean over the Gaussian processes $z^*_t$ and $w^*_t$.  The stochastic Schr\"{o}dinger equation (\ref{NMQSD1}) can be
greatly simplified by using two time-dependent operators $O_1$ and $O_2$ which satisfy consistency conditions $\frac{\delta\psi_t}{\delta
z^*_s}=O_1(t,s,z^*,w^*)\psi_t, \frac{\delta\psi_t}{\delta
w^*_s}=O_2(t,s,z^*,w^*)\psi_t$ \cite{yu2004}. Thus, the NMQSD equation can be written as
\begin{eqnarray} \label{NMQSD2}
\partial_t \psi_t &=& -i H_{\rm eff} \psi_t +
Lz^*_t \psi_t -L^\dag \bar{O}_1(t,z^*,w^*)\psi_t\nonumber\\
&&+L^\dag w^*_t \psi_t -L \bar{O}_2(t,z^*,w^*)\psi_t,
\end{eqnarray}
where
$\bar{O}_i\,\,(i=1,2)$ denotes $
\bar{O}_i(t,z^*,w^*)=
\int^t_0\alpha_i(t,s)O_i(t,s,z^*,w^*)ds,~(i=1,2)$. In a intergal representation, the two correclation functions in a finite temperature can be written as
\begin{eqnarray} \label{correlation}
\alpha_{1}(t,s) &=&\int d\omega\left(\frac{1}{e^{\omega/k_{B}T}-1}+1\right)J(\omega)e^{-i\omega(t-s)},\nonumber \\
\alpha_{2}(t,s) & =&\int d\omega\frac{1}{e^{\omega/k_{B}T}-1}J(\omega)e^{i\omega(t-s)}.
\end{eqnarray}
Their decay, as functions of the time delay $t-s$, defines the ?memory? or correlation time of the environment. 
In the following content, we introduce a Lorentz-Drude spectral density:
$J(\omega) =\frac{\Gamma}{\pi}\frac{\omega}{1+\omega^{2}/\gamma^{2}}$ with a damping rate $\Gamma$.
The two correlation functions are linear at low frequencies, and decays as $1/\omega$ beyond
the cut-off $\gamma$. In the high temperature limit, Eq. (\ref{correlation})
reduces to
\begin{eqnarray}\label{alp}
\alpha_{1}(t,s) &=&k_{B}T\Gamma\varLambda(t,s)+i\Gamma\dot{\varLambda}(t,s),\nonumber\\
\alpha_{2}(t,s)&=&k_{B}T\Gamma\varLambda(t,s),
\end{eqnarray}
where $\varLambda(t,s)=\frac{\gamma}{2}e^{-\gamma|t-s|}$ is an Ornstein-Uhlenbeck correlation function decaying on the environmental memory time $1/\gamma$. Whence we further consider a large cut-off $\gamma$, which means an extremely short memory time, the problem goes back to the Markov case. This is the well-known Caldeira-Leggett (CL) limit $k_{B}T\gg\gamma\gg\omega_{sys}$.

For the purpose of solving the Eq. (\ref{NMQSD2}), we need to determine the operators $O_1$ and $O_2$. Generally, O-operators can be obtained
by invoking a perturbation technique \cite{Yudiosi3,Diosi2}. Thus, in the regime $\Gamma \ll 1$, we get:
\begin{eqnarray}\label{Ooperato1}
\partial_{t}O_{1}(t,s) & =&[-iH_{sys}-L^{\dag}\bar{O}_{1}-L\bar{O}_{2},O_{1}],(L=J_{-})\nonumber\\
\partial_{t}O_{2}(t,s) & =&[-iH_{sys}-L^{\dag}\bar{O}_{1}-L\bar{O}_{2},O_{2}],
\end{eqnarray}
with
\begin{eqnarray}\label{Ooperato2}
O_{1}& =&f_{11}(t,s)J_{-}+f_{12}(t,s)J_{z}J_{-}, \nonumber\\
O_{2} & =&f_{21}(t,s)J_{+}+f_{22}(t,s)J_{+}J_{z},
\end{eqnarray}
where $f_{ij} (i, j=1, 2)$ is a time-dependent coefficient.
Now we substitute Eq. (\ref{Ooperato2}) into Eq. (\ref{Ooperato1}),  ignoring high order terms $J_{+}^{m}J_{z}^{n}(m+n\ge3)$
or $J_{z}^{n}J_{-}^{m}(m+n\ge3)$, the differential equations of the coefficients in the $O_1$ and $O_2$ operators can be derived by
\begin{eqnarray}\label{f11}
\partial_{t}f_{11}&=f_{11}\{ia+ib+J(J-\frac{1}{2})F_{12}\nonumber\\
 & \quad-2F_{21}+(J(J-\frac{1}{2})-2)F_{22}\}\nonumber\\
 & \quad+f_{12}(ibJ-JF_{11}+\frac{J}{2}F_{12}-JF_{21}-\frac{5}{2}JF_{22}),
\end{eqnarray}
\begin{eqnarray}\label{f12}
\partial_{t}f_{12} & =f_{11}\{2ib-2F_{11}+F_{12}-2F_{21}-5F_{22}\}\nonumber\\
 & \quad+f_{12}\{ia+ib+(J(J+1)-3(\frac{3}{2}J-\frac{1}{2}))F_{12}\nonumber\\
 & \quad-2F_{21}+(J(J+1)-2-3(\frac{3}{2}J-\frac{1}{2}))F_{22}\},
\end{eqnarray}

\begin{eqnarray}\label{f21}
\partial_{t}f_{21} & =-f_{21}\{ia+ib+J(J+1)F_{12}\nonumber\\
 & \quad-2F_{21}+(J(J+1)-2)F_{22}\},
 \end{eqnarray}

\begin{eqnarray}\label{f22}
\partial_{t}f_{22} & =-f_{21}\{2ib-2F_{11}+F_{12}-2F_{21}-5F_{22}\}\nonumber\\
 & \quad-f_{22}\{ia+ib+J(J+1)F_{12}\nonumber\\
 & \quad-2F_{21}+(J(J+1)-2)F_{22}\},
\end{eqnarray}
where $F_{ij}=\int_{0}^{t}ds\alpha_i(t, s)f_{ij}(t, s), (i, j=1, 2)$. The initial condition for Eq. (\ref{f11}) to Eq. (\ref{f22}) are $f_{11}(t=s, s)=f_{21}(t=s, s)=1$ and $f_{12}(t=s, s)=f_{22}(t=s, s)=0$.
\begin{center}
\begin{figure}[htp]
 \scalebox{0.50}{\includegraphics{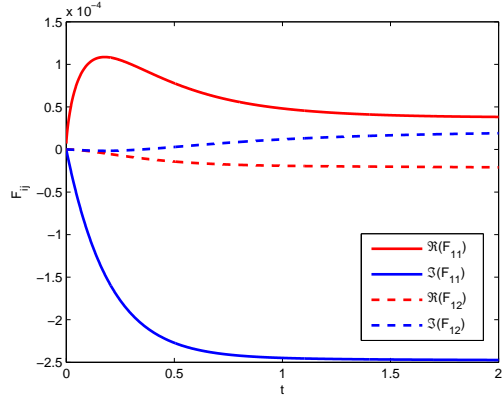}}
 \caption{Time evolution of the coefficients $F_{ij}$ in the O operator. The parameters are $\Gamma=0.0001$; $\gamma=5$; $kT=0.3$ and $N=10$.}
 \end{figure}
\end{center}

In order to derive the master equation, we use the non-Markovian QSD equation (\ref{NMQSD2}) and taking the statistical mean over noises. We obtain \cite{yu2004}
\begin{eqnarray}\label{generalmaster}
\partial_t \rho_t &=& -i[H_{\rm sys},\rho_t]
+[L,\,\, {\mathcal M}\{P_t {\bar O_1}^\dagger(t, z^*,w^*)\}]
-[L^\dagger,\,\,{\mathcal M}\{{\bar
O}_1(t,z^*,w^*) P_t\}]\nonumber\\
&&+[L^\dag,\,\,{\mathcal M}\{P_t{\bar O}_2^\dag(t,z^*,w^*)\}] -
[L,\,\,{\mathcal M}\{\bar{O}_2(t,z^*,w^*)P_t \}].
 \end{eqnarray}
The above equation is the general master equation at finite temperature. Note that this
last result is still not a closed evolution equation for $\rho_t$. Following our previous result in Ref. \cite{yu2004}, the closed master equation can be derived as
\begin{equation}\label{evorho3}
\partial_t \rho = - i[H_{\rm sys},\rho] +
[L,\rho {\bar O}_1^\dagger(t)] + [{\bar
O}_1(t)\rho,L^\dagger]+[L^\dag,\rho {\bar O}_2^\dagger(t)] +
[{\bar O}_2(t)\rho,L],
\end{equation}
with
\begin{eqnarray}
\bar{O}_{1} & =F_{11}(t)J_{-}+F_{12}(t)J_{z}J_{-},\nonumber\\
\bar{O}_{2} & =F_{21}(t)J_{+}+F_{22}(t)J_{+}J_{z},\nonumber\\
\bar{O}_{1}^{\dag} & =F_{11}^{\ast}(t)J_{+}+F_{12}^{\ast}(t)J_{+}J_{z},\nonumber\\
\bar{O}_{2}^{\dag} & =F_{21}^{\ast}(t)J_{-}+F_{22}^{\ast}(t)J_{z}J_{-}.
\end{eqnarray}
In Fig. 1 we give the time dependence of the coefficients $F_{ij}  (i, j=1, 2)$. For the correlation considered in this paper, it shows that the transitions from the non-Markovian to the Markov case through the time evolution. We can clearly see that the non-Markovian environment change the dynamical behavior in the early stage of the time evolution. However, in the Markov regime for the long time evolution, the environment drives the system to the steady state. For a correlation function without steady states, the long-time non-Markovian behavior will be very different \cite{Jingjunetal}.

\section{Spin squeezing}
In order to study spin squeezing in a non-Markovian environment,
we employ the spin squeezing parameter $\xi^2$  \cite{win2,physrep} to ascertain spin squeezing,
which can be defined as
\begin{eqnarray}\label{eq:W0}
\xi^2=\frac{N\langle\Delta J^{2}_{min}\rangle}{\langle J_x\rangle^2},
\end{eqnarray}
where
\begin{eqnarray}
\langle\Delta J^{2}_{min}\rangle=\frac{1}{2}(\langle J^{2}_{y}+J^{2}_{z}\rangle-\sqrt{(\langle J^{2}_{y}-J^{2}_{z}\rangle)^{2}+4\langle J_{y}J_{z}\rangle_{ss}^{2}}),\nonumber
 \end{eqnarray}
 with $\langle J_{y}J_{z}\rangle_{ss}=\langle J_{y}J_{z}+J_{z}J_{y}\rangle/2$.  There is spin squeezing when $\xi^2< 1$.
 \begin{center}
\begin{figure}[htp]\label{gamma}
 \scalebox{0.32}{\includegraphics{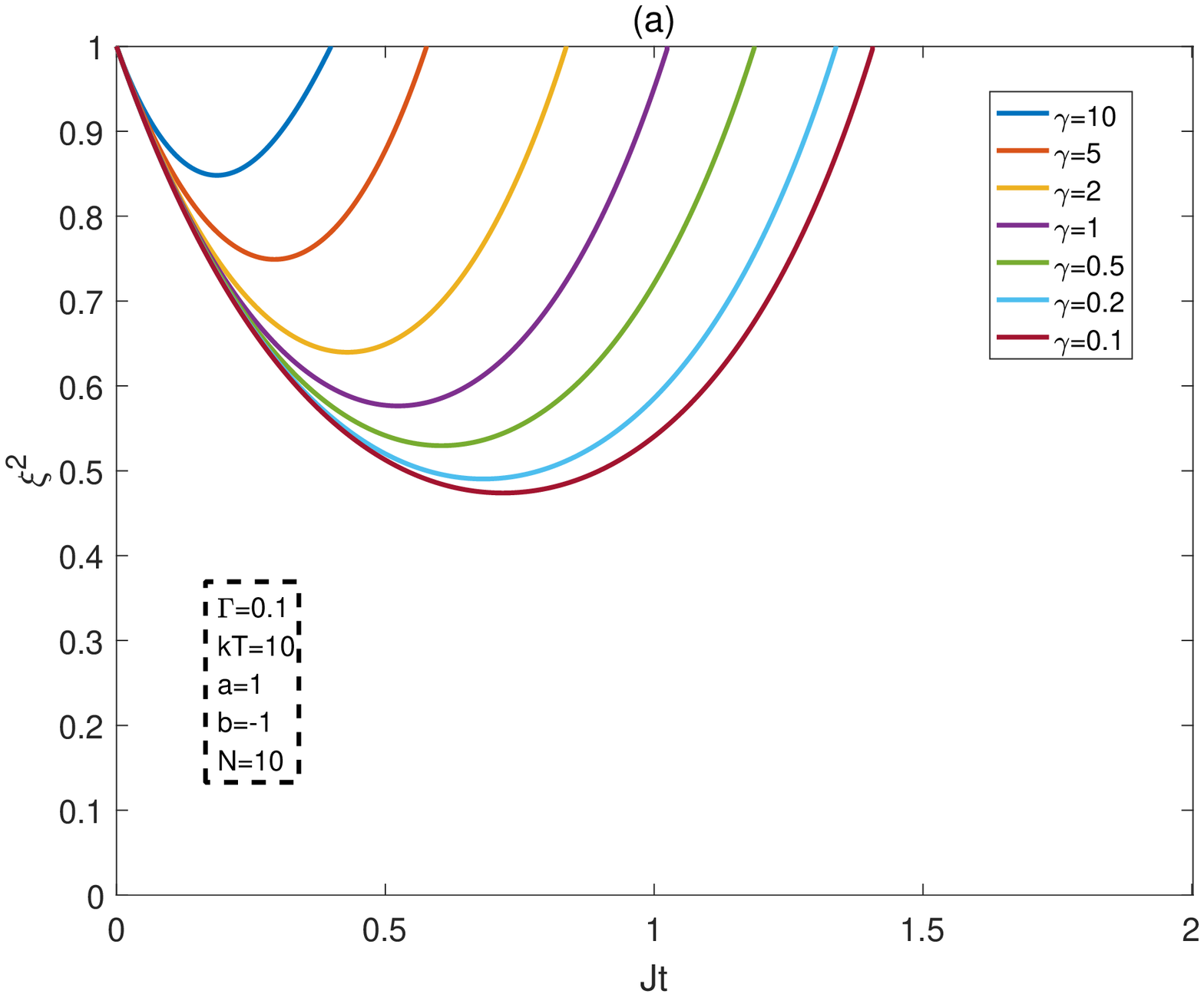}}
 \scalebox{0.32}{\includegraphics{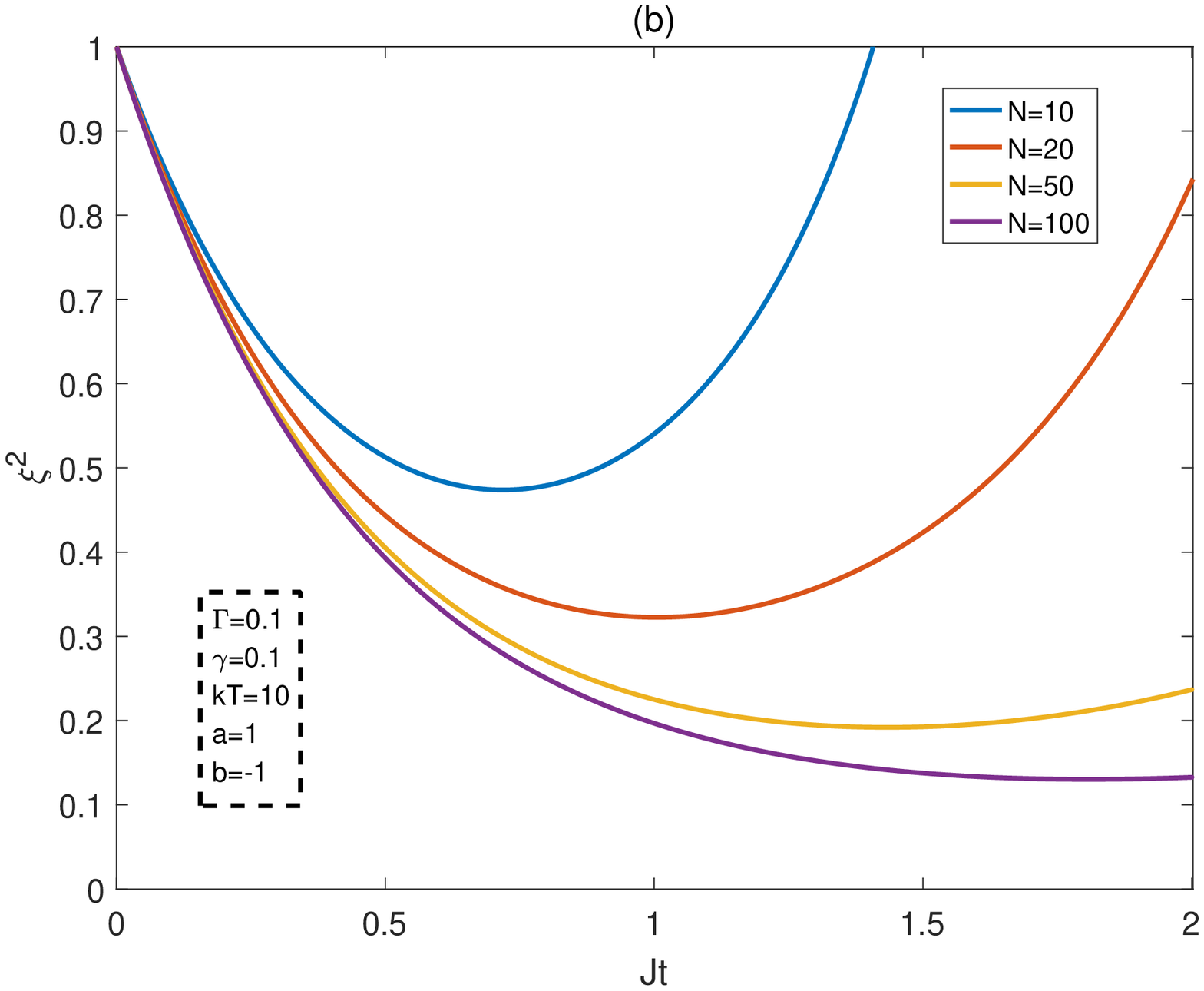}}
\caption{The time evolution of spin squeezing parameter $\xi^2$ with different memory parameter $\gamma$ (Fig. 2a) and with different spin number $N$ (Fig. 2b). The parameters are $\Gamma=0.01$; $kT=10$; $N=20$; $a=1$ and $b=-1$ with different memory parameter $\gamma$: (a)Solid-green line ($\gamma=0.01a$); Dash-red line ($\gamma=0.1a$); Dot-black line ($\gamma=a$) and Dash-dot-blue line ($\gamma=10a$). (b) $\gamma=0.1$ with different spin number $N$: Solid-green line ($N=5$); Dash-red line ($N=20$); Dot-black line ($N=50$) and Dash-dot-blue line ($N=100$).}
 \end{figure}
\end{center}

Fig. 2a shows the evolutions of spin squeezing parameter $\xi^2$ with different memory parameter $\gamma$. For
$\gamma=0.1$ corresponding to a relatively long memory time, the stronger spin squeezing with a long period can be generated, exemplifying
strong non-Markovian effects. The degree of spin squeezing becomes weak with a short period as memory parameter $\gamma$ is increased. Moreover, with the time evolution, the part of exponential decay for non-Markovian process becomes a key element and the spin squeezing disappears finally similar to the Markov decoherence case in the long-time limit.   At $\gamma=10a$ where it nearly approaches Markov regime, it can be seen a very weak spin squeezing only lasting a short-time period. The main reason is that the major decoherence agent is generally the
amplitude damping, the environmental memory is the major determinant
of slowing down the dissipative process due to the back
reaction or information back-flow. Therefore, in order to get the strong and long period spin squeezing, we hope that the dissipative dynamics can temporally reversed and the lost energy or information can come back to the system due to the memory effect. However, for the Markov case, the environment causes the system to decay exponentially into the environment without the information back-flow. Thus, a long memory time not only increases the degrees of the spin squeezing values but also help to achieve the high spin squeezing degrees in shorter time scales as well.

In Fig. 2b,  we plot the evolutions of spin squeezing parameter $\xi^2$ with different spin numbers $N$ for small $\gamma=0.1$ corresponding to a long memory time, and finite temperature. We find that the degree of spin squeezing grows with varying values of $N$, which is agree with the results in Ref.~\cite{benn}, where they have shown that the optimal squeezing degree is proportional to $\sqrt{N}$ ( that is $\xi^2\simeq\frac{2}{\sqrt{J\eta}}$)  in the optimizing $t$ and the detuning $\Delta$.   Thus, relatively large spin numbers not only increase the maximum spin squeezing values but also speed up the whole process.
For a non-Markovian environment, in the case of the experiment realization, the spin number is typically a more convenient parameter that can be effectively controlled. As such, in order to achieve the maximum spin squeezing, one may try to increase the number of spins as large the systems permit.

\begin{center}
\begin{figure}[htp]\label{gamma}
 \scalebox{0.50}{\includegraphics{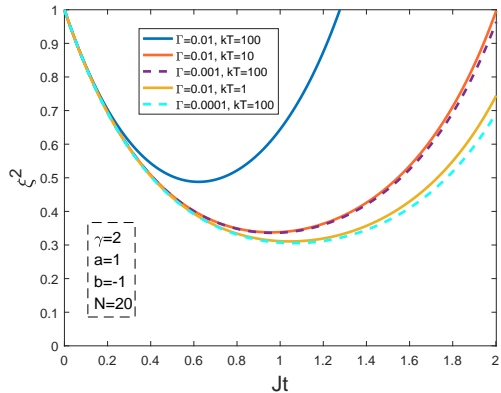}}
\caption{The time evolution of spin squeezing parameter $\xi^2$ with the different damping rate $\Gamma$ and with the different temperature $kT$. The parameters are $\gamma=2$; $N=20$; $a=1$ and $b=-1$.}
 \end{figure}
\end{center}

Moreover, in Fig. 3,  we consider the effects of the environment decay rate $\Gamma$ and the temperature $T$ on the spin squeezing in the non-Markovian environment.  For this purpose, the environment decay rates are chosen as $\Gamma=0.0001$ (blue line), $\Gamma=0.001$ (purple line) and $\Gamma=0.01$ (green line), respectively,  we find that the time range of spin squeezing may be prolonged as the decay rate $\Gamma$ decreases. The reason is obvious as the weaker decay rates will maintain system's dynamics longer.  On the other hand, choosing the different environment temperatures for $kT=1$ (yellow line), $kT=10$ (orange line) and $kT=100$ (blue line), it shows that the time range of spin squeezing can be narrowed
as the temperature $T$ increases. As shown above, the lower environment decay rate or environment temperature will lead a wider time range of the spin squeezing.

\begin{center}
\begin{figure}[htp]
 \scalebox{0.37}{\includegraphics{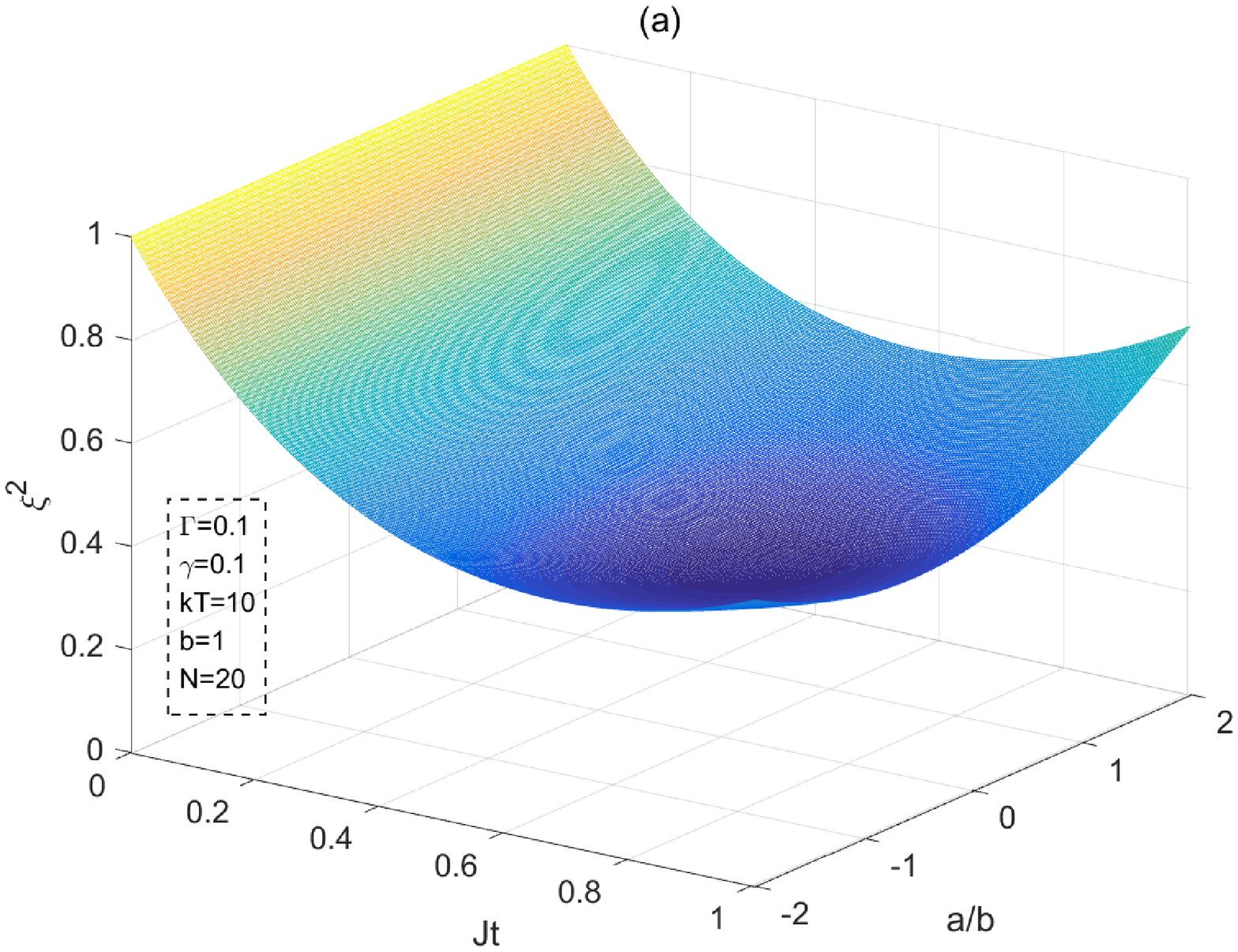}}
 \scalebox{0.37}{\includegraphics{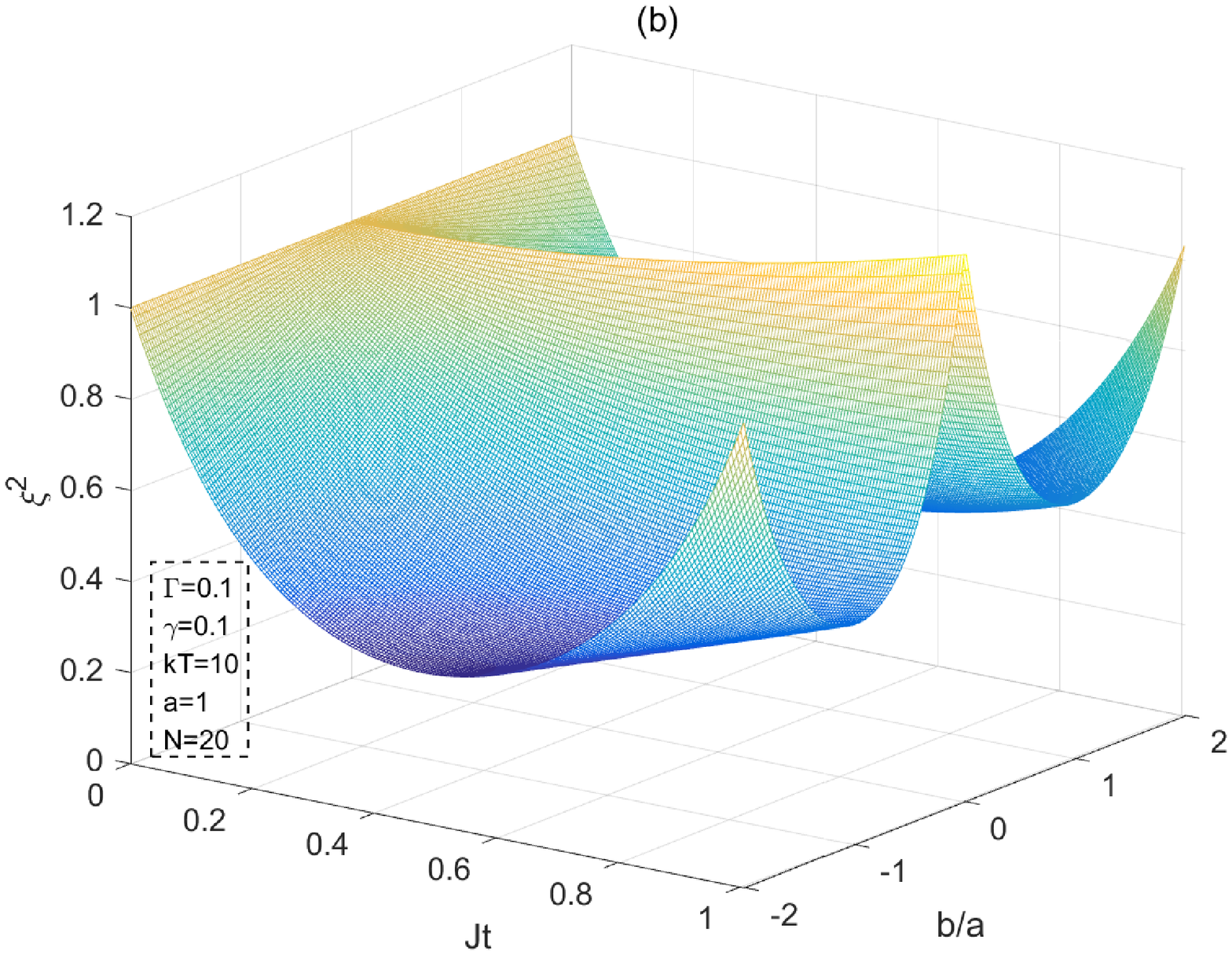}}
\caption{The time evolution of spin squeezing parameter $\xi^2$ with different ratio $a/b$ (Fig. 4a) and different ratio $b/a$ (Fig. 4b). The parameters are $\Gamma=0.01$; $\gamma=2$; $kT=10$ and $N=20$.}
 \end{figure}
\end{center}

Apart from above spin numbers, another important feature
of the environment is dictated by the parameters $a$ and $b$, which is shown to be important in spin squeezing generation. In Fig. 4, we plot the evolution  of the spin squeezing parameter $\xi^2$ with different ratio $a/b$ and $b/a$.
It is obvious that spin squeezing parameter $\xi^2$ is less than 1 for the time interval from 0 to $Jt=1$ in the low ratio  of $a/b$ and $b/a$ (see also Fig. 4). Fig. 4a shows that the minimum spin squeezing occurs at the point that is close to the ratio $a/b=0$ and the spin squeezing parameter $\xi^2$ is a symmetrical distribution on the both sides of $a/b=0$ with the time evolution. As the ratio $a/b$ increases, we can see, with time development, that the spin squeezing parameter $\xi^2$ eventually returns to 1 which means that the spin squeezing is to disappear.  This interesting observation may be explained by Eq. (3).  In fact, 
the first term in Eq. (3) represents the nonlinear interaction for the one-axis-twisting (OAT) type spin squeezing, here $b$ is the spins interaction parameter. When $a\sim0$, it becomes OTA type Hamiltonian in the form of $H=bJ_z^2$, which can generate OTA-type  spin squeezing as confirmed by the recent spin squeezing experiment through the state-selective collisions  \cite{est,gro,rie}.

From Fig.~4b, if we increase the ratio $b/a$, as the time evolves, the spin squeezing parameter $\xi^2$ will exhibit two extreme points located symmetrically around $b/a=0$.
However, we find that the spin squeezing parameter $\xi^2$ can be close to 1 when $b/a\sim0$ for all time points. The reason is that, when the nonlinear term $bJ_z^2$ is absent, the linear term $aJ_z$ alone can not generate the spin squeezing.   For $b/a=0$ the two levels of the spins are almost equally populated and the spins may form a standing wave with pronounced interference fringes \cite{kolar,opa}.   Therefor, these spins may be effectively compressed to the half of the occupied volume if these spins  all circulate in the same direction ($J_z\approx\pm N/2$) without interference fringes \cite{kolar},  here small $b/a$ means lower interaction energy.  As an interesting result, from Fig. 4b,  we also  find the spin squeezing  occurs when  $b=0$.  Moreover, it also shows that time range of spin squeezing will increase first and then decrease as the ratio $b/a$ increases from 0 to 2. It is shown that spin squeezing generation begins when the ratio $b/a$ is about $0.3$.

\section{Discussion and Conclusion}
Generation of spin squeezing with the LMG-type Hamiltonian may be realized in different physical settings such as  Bose-Einstein condensation (BEC) systems \cite{BEC1,BEC2,BEC3,BEC4,BEC5,BEC6,BEC7,BEC8,BEC9}. The major advantage of producing spin squeezing in the BEC systems is  the possible realization of the strong
atom-atom interactions, which can induce the nonlinearity and squeezing \cite{physrep}. It has shown that, in the single-mode approximation, the Hamiltonian described in Eq.~(\ref{ham}) of $N$ atoms with two internal states $|a\rangle$ and $|b\rangle$, may be realized \cite{rensen}. A similar Hamiltonian for creating spin squeezing may be also realized with nitrogen-vacancy centers in diamond \cite{benn}. It is demonstrated that the direct spin-phonon
coupling in diamond can be used to generate spin squeezed states of an NV ensemble embedded in the Markov environment. Finally, we note that Hamiltonian defined in Eq.~(\ref{ham})
has been studied in the context of nuclear spins in a quantum dot system~\cite{rud}, where they have achieved spin squeezing by using the same LMG Hamiltonian. We emphasize that these physical settings may also be used to realize the non-Markovian spin squeezing after necessary modifications. For example, in the case of BEC system, it is possible to achieve the strong coupling regimes for atom-light interaction, therefore, the non-Markovian features discussed here may be studied in these types of systems.

In conclusion, we study the LMG model to generate spin squeezing in the presence of non-Markovian environments at finite temperature. The model is composed of $N$ identical spin-1/2 particles interacting with the identical bosonic heat baths. We have derived the finite-temperature non-Markovian master equations by using NMQSD method. In non-Markovian regimes, we numerically investigate the dependence of spin squeezing on several relevant physical parameters including the environment memory times, temperatures, and decay rates. It is shown that  the finite memory-time scales or large spin numbers can increase both the maximum spin squeezing values  and the squeezing time periods. Moreover, we examine the spin squeezing of the different LMG models corresponding to the different choices of the parameters $a$ and $b$ in Hamiltonian Eq.~(\ref{ham}). The results show that, for some parameter domains of $a$ and $b$, the relative strong and long-lived spin squeezing may be obtained.

\appendix
\section{Expectation of collective operators with time evolution}
In this appendix, we outline the mean values that enter the variance
of Eq.~\eqref{eq:W0}. As is known, for an operator $\hat{A}$ ,
the expectation with time evolution can be expressed by means of the
master equation as $d\langle\hat{A}\rangle/dt=\text{tr}[\hat{A}\dot{\rho}]$.
Using Eq.~\eqref{evorho3} we can easily obtain the following equations

\begin{eqnarray}
\frac{d}{dt}\langle A\rangle & =-i\langle[A,H_{sys}]\rangle+\{\langle\bar{O}_{1}^{\dagger}[A,L]\rangle+\langle[L^{\dagger},A]\bar{O}_{1}\rangle+\langle\bar{O}_{2}^{\dagger}[A,L^{\dagger}]\rangle+\langle[L,A]\bar{O}_{2}\rangle\}.
\end{eqnarray}
\begin{eqnarray}
\frac{d}{dt}\langle J_{-}\rangle = & -i(a+b)\langle J_{-}\rangle-2ib\langle J_{z}J_{-}\rangle\nonumber\\
 & \quad+\{2F_{11}(\langle J_{z}J_{-}\rangle)-2F_{21}^{*}(\langle J_{z}J_{-}\rangle+\langle J_{-}\rangle)-2F_{22}^{*}(\langle J_{z}J_{-}\rangle+\langle J_{z}^{2}J_{-}\rangle)\} \\
\frac{d}{dt}\langle J_{z}\rangle = & 2\Re\{-F_{11}^{*}(J(J+1)-\langle J_{z}^{2}\rangle+\langle J_{z}\rangle)-F_{12}^{*}(-J(J+1)+(J(J+1)-1)\langle J_{z}\rangle+2\langle J_{z}^{2}\rangle-\langle J_{z}^{3}\rangle)\nonumber\\
 & \quad+F_{21}^{*}(J(J+1)-\langle J_{z}^{2}\rangle-\langle J_{z}\rangle)+F_{22}^{*}(J(J+1)+(J(J+1)-1)\langle J_{z}\rangle-2\langle J_{z}^{2}\rangle-\langle J_{z}^{3}\rangle)\}\\
\frac{d}{dt}\langle J_{-}^{2}\rangle = & -i(2a+4b)\langle J_{-}^{2}\rangle-4ib\langle J_{z}J_{-}^{2}\rangle +F_{11}(4\langle J_{z}J_{-}^{2}\rangle+2\langle J_{-}^{2}\rangle)+F_{12}(4\langle J_{z}^{2}J_{-}^{2}\rangle\nonumber\\
 &+6\langle J_{z}J_{-}^{2}\rangle+2\langle J_{-}^{2}\rangle) -F_{21}^{*}(4\langle J_{z}J_{-}^{2}\rangle+6\langle J_{-}^{2}\rangle) -F_{22}^{*}(4\langle J_{z}^{2}J_{-}^{2}\rangle+6\langle J_{z}J_{-}^{2}\rangle)\\
\frac{d}{dt}\langle J_{z}^{2}\rangle = & 2\Re\{-F_{11}^{*}(-J(J+1)+(2J(J+1)-1)\langle J_{z}\rangle+3\langle J_{z}^{2}\rangle-2\langle J_{z}^{3}\rangle)\nonumber\\
 & \quad-F_{12}^{*}(J(J+1)-(3J(J+1)-1)\langle J_{z}\rangle+(2J(J+1)-4)\langle J_{z}^{2}\rangle+5\langle J_{z}^{3}\rangle-2\langle J_{z}^{4}\rangle)\nonumber\\
 & \quad+F_{21}^{*}(J(J+1)+(2J(J+1)-1)\langle J_{z}\rangle-3\langle J_{z}^{2}\rangle-2\langle J_{z}^{3}\rangle)\nonumber\\
 & \quad+F_{22}^{*}(J(J+1)\langle J_{z}\rangle+(2J(J+1)-1)\langle J_{z}^{2}\rangle-3\langle J_{z}^{3}\rangle-2\langle J_{z}^{4}\rangle)\}\\
\frac{d}{dt}\langle J_{z}J_{-}\rangle = & -ia\langle J_{z}J_{-}\rangle-ib(\langle J_{z}J_{-}\rangle+2\langle J_{z}^{2}J_{-}\rangle)+\{-F_{11}(J(J+1)\langle J_{-}\rangle+\langle J_{z}J_{-}\rangle-3\langle J_{z}^{2}J_{-}\rangle)\nonumber\\
 & -F_{11}^{*}(J(J+1)\langle J_{-}\rangle+\langle J_{z}J_{-}\rangle-\langle J_{z}^{2}J_{-}\rangle)-F_{12}(J(J+1)\langle J_{z}J_{-}\rangle+\langle J_{z}^{2}J_{-}\rangle-3\langle J_{z}^{3}J_{-}\rangle)\nonumber\\
 & -F_{12}^{*}(-J(J+1)\langle J_{-}\rangle+(J(J+1)-1)\langle J_{z}J_{-}\rangle+2\langle J_{z}^{2}J_{-}\rangle-\langle J_{z}^{3}J_{-}\rangle)+F_{21}((J(J+1)-2)\langle J_{-}\rangle\\
 &-3\langle J_{z}J_{-}\rangle-\langle J_{z}^{2}J_{-}\rangle) +F_{21}^{*}((J(J+1)-2)\langle J_{-}\rangle-5\langle J_{z}J_{-}\rangle-3\langle J_{z}^{2}J_{-}\rangle)\nonumber\\
 & +F_{22}((J(J+1)-2)\langle J_{-}\rangle+(J(J+1)-5)\langle J_{z}J_{-}\rangle-4\langle J_{z}^{2}J_{-}\rangle-\langle J_{z}^{3}J_{-}\rangle)\\
 & +F_{22}^{*}((J(J+1)-2)\langle J_{z}J_{-}\rangle-5\langle J_{z}^{2}J_{-}\rangle-3\langle J_{z}^{3}J_{-}\rangle)\}
\end{eqnarray}
\section*{Acknowledgements}

The project was supported by NSFC (Grants No.11164029) and Distinguished Young Scholars Breeding
fund of Inner Mongolia (Grant No. 2017JQ08).   


\end{document}